\documentclass[12pt,a4paper,english,nofootinbib]{revtex4}
\usepackage{lmodern}
\usepackage{lmodern}

\usepackage[T1]{fontenc}
\usepackage[latin9]{inputenc}
\usepackage{babel}
\usepackage{amsmath}
\usepackage{amssymb}
\usepackage{graphicx}
\usepackage{esint}
\usepackage[unicode=true,pdfusetitle,
 bookmarks=true,bookmarksnumbered=false,bookmarksopen=false,
 breaklinks=false,pdfborder={0 0 1},backref=false,colorlinks=false]
 {hyperref}

\makeatletter

\newcommand{\lyxmathsym}[1]{\ifmmode\begingroup\def\b@ld{bold}
  \text{\ifx\math@version\b@ld\bfseries\fi#1}\endgroup\else#1\fi}

\@ifundefined{textcolor}{}
{%
 \definecolor{BLACK}{gray}{0}
 \definecolor{WHITE}{gray}{1}
 \definecolor{RED}{rgb}{1,0,0}
 \definecolor{GREEN}{rgb}{0,1,0}
 \definecolor{BLUE}{rgb}{0,0,1}
 \definecolor{CYAN}{cmyk}{1,0,0,0}
 \definecolor{MAGENTA}{cmyk}{0,1,0,0}
 \definecolor{YELLOW}{cmyk}{0,0,1,0}
}

\usepackage{babel}
\usepackage{babel}
\usepackage{babel}
\usepackage{babel}
\usepackage{babel}
\usepackage{babel}
\usepackage{babel}
\usepackage{babel}
\usepackage{babel}
\usepackage{babel}
\usepackage{babel}
\usepackage{babel}
\usepackage{babel}
\usepackage{babel}
\usepackage{babel}
\usepackage{babel}
\usepackage{babel}
\usepackage{babel}
\usepackage{babel}
\usepackage{babel}
\usepackage{babel}
\usepackage{babel}

\usepackage{babel}
\def\b{\begin{equation}}
\def\e{\end{equation}}

\@ifundefined{textcolor}{}{%
 \definecolor{BLACK}{gray}{0}
 \definecolor{WHITE}{gray}{1}
 \definecolor{RED}{rgb}{1,0,0}
 \definecolor{GREEN}{rgb}{0,1,0}
 \definecolor{BLUE}{rgb}{0,0,1}
 \definecolor{CYAN}{cmyk}{1,0,0,0}
 \definecolor{MAGENTA}{cmyk}{0,1,0,0}
 \definecolor{YELLOW}{cmyk}{0,0,1,0}
 }

\usepackage{latexsym}\usepackage{bm}

\makeatother

\begin{document}
\title{{\normalsize{}{}{}{}{}{}{}{}{}{}{}Approximate analytical
description of apparent horizons for initial data with momentum and
spin}}
\author{{\normalsize{}{}{}{}{}{}{}{}{}{}{}{}{}{}{}{}{}{}Emel
Altas}}
\email{emelaltas@kmu.edu.tr}

\affiliation{Department of Physics,\\
 Karamanoglu Mehmetbey University, 70100, Karaman, Turkey}
\author{{\normalsize{}{}{}{}{}{}{}{}{}{}{}{}{}{}{}{}{}{}Bayram
Tekin}}
\email{btekin@metu.edu.tr}

\affiliation{Department of Physics,\\
 Middle East Technical University, 06800, Ankara, Turkey}
\date{{\normalsize{}{}{}{}{}{}{}{}{}{}{}{{}{}{}{}\today}}}

\maketitle
We construct
analytical initial data for a slowly moving and rotating black hole
for generic orientations of the linear momentum and the spin.  We solve the Hamiltonian constraint approximately and work out the properties of the apparent horizon and show the dependence of its shape on the angle between the spin and the linear momentum. In particular a dimple, whose location depends on the mentioned angle, arises on the 2-sphere geometry of the apparent horizon. We exclusively work in the case of conformally flat initial metrics. 

\section{{\normalsize{}{}Introduction}}

Since the first observation of black hole merger \cite{merger}, there
have been many observations of merger of compact objects via gravitational
waves. The gravitational waves produced by these mergers are consistent
with the numerical solutions of the field equations of General Relativity.
Besides the highly accurate numerical results, it always pays to have
approximate solutions of relativistic gravitating systems. Here we
give an approximate analytical description of a self-gravitating system
that has a conserved total energy, total spin and a linear momentum
in an asymptotically flat spacetime. The initial configuration is
expected to evolve and settle to a single rotating black hole after emitting
some gravitational radiation.

The problem was studied in \cite{Gleiser} in the case of vanishing
linear momentum but with a nonzero spin; and in \cite{Dennison-Baumgarte}
in the case of vanishing spin with a nonzero linear momentum.  See a remarkable exposition in \cite{kitap}. Here
we assume both of these quantities to be nonzero and pointing arbitrarily
in three dimensional space. It will turn out that the shape of the
apparent horizon depends on the angle between the linear momentum
and the spin: even though at the next to leading order, the magnitude
of the spin does not appear in the shape of the apparent horizon,
its direction does. On the other hand, the shape of the apparent horizon depends
on the magnitude of the linear momentum at the first order. The area of the apparent horizon does not depend on the angle between the spin and linear momentum.
 We also observe that a dimple arises on the 2-sphere geometry of the apparent horizon. 

The layout of the paper is as follows. In the next section, we discuss
briefly the constraint equations in General Relativity and present
the Bowen-York method \cite{BY} in finding solutions to the initial value
problem. In section $\lyxmathsym{\mbox{III}}$, we give the approximate
solution of the Hamiltonian constraint for a slowly rotating and moving black hole. In section $\text{\mbox{IV}}$, 
we compute the position of the apparent horizon as a function of the angle between the spin and the linear momentum.

\section{{\normalsize{}{}{}{}{}{}{}{}{}{}{}{}{}{}{}{}{}{}Initial
data for a Black hole with momentum and spin}}

Assuming the usual ADM split of the metric \cite{ADM} 
\begin{equation}
ds^{2}=(N_{i}N^{i}-N^{2})dt^{2}+2N_{i}dtdx^{i}+\gamma_{ij}dx^{i}dx^{j},\hskip1cmi,j\in(1,2,3),\label{ADMdecompositionofmetric}
\end{equation}
the Einstein equations in vacuum without a cosmological constant split
into constraints and the evolution equations. The constraint equations
are given as 
\begin{eqnarray}
 &  & -^{\Sigma}R-K^{2}+K_{ij}K^{ij}=0,\nonumber \\
 &  & -2D_{k}K_{i}^{k}+2D_{i}K=0,~~~~~~~~~~~~~~\label{Einstein_c}
\end{eqnarray}
where $\Sigma$ is the Cauchy surface; 
$K_{ij}=K_{ij}(t,x^{k})$ is its extrinsic curvature defined as 
\begin{equation}
K_{ij}=\frac{1}{2N}\Big(\dot{\gamma}_{ij}-D_{i}N_{j}-D_{j}N_{i}\Big),\hskip1cm\dot{\gamma}_{ij}=\frac{\partial}{\partial t}\gamma_{ij},
\end{equation}
with the trace $K:=\gamma^{ij}K_{ij}$; and $D_{i}\gamma_{kl}=0$.
For further details of the construction, including the evolution equations
which we do not depict here, see the Appendix of \cite{our_dain_paper}.
Following Bowen-York \cite{BY}, let us assume that $\Sigma$ is conformally
flat with the metric

\begin{equation}
\gamma_{ij}=\psi^{4}f_{ij},\hskip1cm\psi>0,
\end{equation}
with $f_{ij}$ denoting the flat metric in some generic coordinates.
One also sets the extrinsic curvature of the hypersurface to be given
as $K_{ij}=\psi^{-2}\hat{K}_{ij}$. Furthermore, we assume that $\Sigma$
is a maximally embedded hypersurface in the spacetime such that the trace of the extrinsic curvature vanishes~\footnote{For physically relevant decay conditions in the case of asymptotically flat initial data, we refer the reader to Section III C of  \cite{our_dain_paper} where a slightly extended discussion is compiled.}
\begin{equation}
K=0.
\end{equation}
Under these conditions the Hamiltonian constraint reduces to a nonlinear elliptic equation
\begin{equation}
\hat{D}_{i}\hat{D}^{i}\psi=-\frac{1}{8}\psi^{-7}\hat{K}_{ij}^{2},\label{elliptic0}
\end{equation}
and the momentum constraint reduces to 
\begin{equation}
\hat{D}^{i}\hat{K}_{ij}=0\label{trans},
\end{equation}
with $\hat{D}_{i}f_{jk}=0$. The momentum constraint equations can
be solved easily, following \cite{BY}, let us choose the
$6$ parameter solution
\begin{equation}
\hat{K}_{ij}=\frac{3}{2r^{2}}\Big(p_{i}n_{j}+p_{j}n_{i}+(n_{i}n_{j}-f_{ij})p\cdot n\Big)+\frac{3}{r^{3}}{\cal {J}}^{l}n^{k}\Big(\varepsilon_{kil}n_{j}+\varepsilon_{kjl}n_{i}\Big),\label{Bowen-York extrinsic curvature a=00003D00003D00003D00003D0}
\end{equation}
where $n^{i}$ is the unit normal on a sphere of radius $r$. For other solutions, see \cite{Beig}.
Assuming the following asymptotic behavior for the conformal factor
\begin{equation}
\psi(r)=1+\frac{E}{2r}+{\mathcal{O}}(1/r^{2}),\label{asymp1}
\end{equation}
one can easily show that (see \cite{Altas}) the $p^{i}$ in the solution
(\ref{Bowen-York extrinsic curvature a=00003D00003D00003D00003D0})
corresponds to the total conserved linear momentum via 
\begin{equation}
P_{i}=\frac{1}{8\pi}\int_{S_{\infty}^{2}}dS\,n^{j}\,K_{ij}=\frac{1}{8\pi}\int_{S_{\infty}^{2}}dS\,n^{j}\,\hat{K}_{ij}.\label{mom}
\end{equation}
Similarly one can show that $J_{i}$ corresponds to the total conserved angular
momentum expressed in terms of the coordinates and the extrinsic curvature
as 
\begin{equation}
J_{i}=\frac{1}{16\pi}\varepsilon_{ijk}\int_{S_{\infty}^{2}}dS\,n_{l}\,\Big(x^{j}K^{kl}-x^{k}K^{jl}\Big)=\frac{1}{16\pi}\varepsilon_{ijk}\int_{S_{\infty}^{2}}dS\,n_{l}\,\Big(x^{j}\hat{K}^{kl}-x^{k}\hat{K}^{jl}\Big).\label{dad}
\end{equation}
Finally the ADM energy 
\begin{equation}
E_{ADM}=\frac{1}{16\pi}\int_{S_{\infty}^{2}}dS\,n_{i}\,\Big(\partial_{j}h^{ij}-\partial_{i}h_{j}^{j}\Big),
\end{equation}
becomes 
\begin{equation}
E_{ADM}=-\frac{1}{2\pi}\int_{S_{\infty}^{2}}dS\,n^{i}\,\partial_{i}\psi,
\end{equation}
and so using the asymptotic form (\ref{asymp1}) one finds $E_{ADM}=E$. This has been a brief description of the solution of the momentum
constraints. Now, the important task is to solve
the Hamiltonian constraint, which as we noted, is a nonlinear elliptic equation
and thus, generically, it can only be solved numerically. But in the next
section, we shall give an approximate solution for small momentum and
small rotation.

\section{{\normalsize{}{}Initial data with small momentum and small spin}}

Computation of $\hat{K}_{ij}\hat{K}^{ij}$ (from equation (\ref{Bowen-York extrinsic curvature a=00003D00003D00003D00003D0}))
yields

\begin{equation}
\hat{K}_{ij}\hat{K}^{ij}=\frac{9}{2r^{4}}\left(p^{2}+2(\vec{p}\cdot\vec{n})^{2}\right)+\frac{18}{r^{5}}\left(\vec{J}\times\vec{n}\right)\cdotp\vec{p}+\frac{18}{r^{6}}\left(\vec{J}\times\vec{n}\right)\cdotp\left(\vec{J}\times\vec{n}\right).
\end{equation}
Without loss of generality, let us assume that the direction of the
spin is the $\hat{k}$ direction, namely 
\begin{equation}
\vec{J}=J\hat{k},
\end{equation}
and $\vec{p}$ is lying in the $xz$ plane and given as 
\begin{equation}
\vec{p}=p\sin\theta_{0}\hat{i}+p\cos\theta_{0}\hat{k}
\end{equation}
with $\theta_{0}$ a fixed angle. To simplify the notation of the following
discussion, let us denote 
\begin{equation}
c_{1}:=\sin\theta_{0},~~~~~~~~c_{2}:=\cos\theta_{0}.
\end{equation}
The Hamiltonian constraint, after these conventions becomes 
\begin{equation}
\hat{D}_{i}\hat{D}^{i}\psi=\psi^{-7}\left(\frac{9Jp}{4r^{5}}c_{1}\sin\theta\sin\phi-\frac{9J^{2}}{4r^{6}}\sin^{2}\theta-\frac{9p^{2}}{16r^{4}}(1+2(c_{1}\sin\theta\cos\phi+c_{2}\cos\theta)^{2})\right).\label{hamiltoian constraint}
\end{equation}
As it clear from the right-hand side, the correct perturbative expansion
in terms of the momentum and spin reads

\begin{equation}
\psi(r,\theta,\phi):=\psi^{(0)}+J^{2}\psi^{(J)}+p^{2}\psi^{(p)}+Jp\psi^{(Jp)}+\mathcal{O}(p^{4},J^{4},p^{2}J^{2}),\label{expansionof psi}
\end{equation}
where the functions on the right-hand side depend on all coordinates
$(r,\theta,\phi)$. At the lowest order, one has 
\begin{equation}
\hat{D}_{i}\hat{D}^{i}\psi^{(0)}=0.
\end{equation}
To proceed, let us discuss the boundary conditions that we shall employ.
Following \cite{Brandt} and \cite{Dennison-Baumgarte}, we chose the following boundary
conditions

\begin{equation}
\lim_{r\rightarrow\infty}\psi(r)=1,~~~~~~~~~\psi(r)>0,\label{boundarycondition1}
\end{equation}
and 
\begin{equation}
\lim_{r\rightarrow\ 0}\psi(r)=\psi^{(0)}.\label{boundarycondition2}
\end{equation}
At the lowest order, the solution satisfying these boundary conditions
reads 
\begin{equation}
\psi^{(0)}=1+\frac{a}{r}.
\end{equation}
Inserting (\ref{expansionof psi}) into (\ref{hamiltoian constraint}),
one arrives at three linear partial differential equations to be solved:
\begin{equation}
\hat{D}_{i}\hat{D}^{i}\psi^{(J)}=-\frac{9}{4}\sin^{2}\theta\frac{r}{(r+a)^{7}}, \hskip 0.5 cm
\hat{D}_{i}\hat{D}^{i}\psi^{(Jp)}=\frac{9}{4}c_{1}\sin\theta\sin\phi\frac{r^{2}}{(r+a)^{7}},\label{Jpequation}
\end{equation}
and also 
\begin{equation}
\hat{D}_{i}\hat{D}^{i}\psi^{(p)}=-\frac{9}{16}\left(1+2(c_{1}\sin\theta\cos\phi+c_{2}\cos\theta)^{2}\right)\frac{r^{3}}{(r+a)^{7}}.\label{pequation}
\end{equation}
In finding the solutions to these equations, we will need the following
spherical harmonics : 
\begin{equation}
Y_{0}^{0}(\theta,\phi)=\frac{1}{\sqrt{4\pi}},~~~~~~~~~~~~~~~Y_{1}^{0}(\theta,\phi)=\sqrt{\frac{3}{4\pi}}\cos\theta,~~~~~~Y_{2}^{0}(\theta,\phi)=\sqrt{\frac{5}{16\pi}}(3\cos^{2}\theta-1),
\end{equation}
\[
Y_{1}^{-1}(\theta,\phi)=\sqrt{\frac{3}{4\pi}}\sin\theta\sin\phi,~~~~Y_{2}^{1}(\theta,\phi)=\sqrt{\frac{15}{4\pi}}\sin\theta\cos\theta\cos\phi,~~~~Y_{1}^{1}(\theta,\phi)=\sqrt{\frac{3}{4\pi}}\sin\theta\cos\phi.
\]
Then ansatz for $\psi^{(J)}$ can be taken as 
\begin{equation}
\psi^{(J)}(r,\theta,\phi)=\psi_{0}^{(J)}(r)Y_{0}^{0}(\theta,\phi)+\psi_{1}^{(J)}(r)Y_{2}^{0}(\theta,\phi),
\end{equation}
which upon insertion to the first equation of (\ref{Jpequation}) yields two ordinary differential
equations equations 
\begin{equation}
\frac{d}{dr}\bigl(r^{2}\frac{d\psi_{0}^{(J)}(r)}{dr}\bigr)=-3\sqrt{\pi}\frac{r^{3}}{(r+a)^{7}},\hskip 0.5 cm
\frac{d}{dr}\bigl(r^{2}\frac{d\psi_{1}^{(J)}(r)}{dr}\bigr)-6\psi_{1}^{(J)}(r)=3\sqrt{\frac{\pi}{5}}\frac{r^{3}}{(r+a)^{7}}.
\end{equation}
The solution obeying the boundary conditions (\ref{boundarycondition1}) reads
\begin{equation}
\psi^{(J)}(r,\theta,\phi)=\frac{\left(a^{4}+5a^{3}r+10a^{2}r^{2}+5ar^{3}+r^{4}\right)}{40a^{3}(a+r)^{5}}-\frac{r^{2}}{40a(a+r)^{5}}(3\cos^{2}\theta-1).
\end{equation}
Similarly setting 
\begin{equation}
\psi^{(Jp)}(r,\theta,\phi)=\psi_{0}^{(Jp)}(r)Y_{0}^{0}(\theta,\phi)+\psi_{1}^{(Jp)}(r)Y_{1}^{-1}(\theta,\phi)
\end{equation}
in  the second equation of (\ref{Jpequation}), one finds that $\psi_{0}^{(Jp)}(r)=0$ satisfies
the boundary conditions; and the $\psi_{1}^{(Jp)}(r)$ piece satisfies
\begin{equation}
\frac{d}{dr}\bigl(r^{2}\frac{d\psi_{1}^{(Jp)}(r)}{dr}\bigr)-2\psi_{1}^{(Jp)}(r)=\frac{3\sqrt{3\pi}}{2}c_{1}\frac{r^{4}}{(r+a)^{7}}\label{kolay2}
\end{equation}
of which the solution can be found and one has 
\begin{equation}
\psi^{(Jp)}(r,\theta,\phi)=-\frac{c_{1}r\left(a^{2}+5ar+10r^{2}\right)}{80a(a+r)^{5}}\sin\theta\sin\phi.
\end{equation}
Finally, let us do the $\psi^{(p)}(r,\theta,\phi)$ part which is slightly more
complicated. One sets 
\begin{equation}
\psi^{(p)}=\psi_{0}^{(p)}(r)Y_{0}^{0}(\theta,\phi)+\psi_{1}^{(p)}(r)Y_{1}^{1}(\theta,\phi)^{2}+\psi_{2}^{(p)}(r)Y_{2}^{1}(\theta,\phi)+\psi_{3}^{(p)}(r)Y_{1}^{0}(\theta,\phi)^{2}
\end{equation}
to arrive at four equations, two of which are 
\begin{equation}
\frac{d}{dr}\bigl(r^{2}\frac{d\psi_{0}^{(p)}}{dr}\bigr)+\frac{3}{\sqrt{\pi}}(\psi_{1}^{(p)}+\psi_{3}^{(p)})=-\frac{9}{8}\sqrt{\pi}\frac{r^{5}}{(r+a)^{7}},
\end{equation}
and 
\begin{equation}
\frac{d}{dr}\bigl(r^{2}\frac{d\psi_{1}^{(p)}}{dr}\bigr)-6\psi_{1}^{(p)}=-\frac{3}{2}\pi c_{1}^{2}\frac{r^{5}}{(r+a)^{7}}.
\label{psi1}
\end{equation}
$\psi_{2}^{(p)}(r)$ equation can be obtained from (\ref{psi1}) with
the replacement $c_{1}^{2}\rightarrow\sqrt{\frac{3}{5\pi}}c_{1}c_{2}$
and $\psi_{3}^{(p)}(r)$ equation can be obtained from (\ref{psi1})
via $c_{1}^{2}\rightarrow c_{2}^{2}$.  
The solutions read, respectively, as follows 
\begin{eqnarray}
\psi_{0}^{(p)}(r) & = & -\frac{\sqrt{\pi}\left(84a^{6}+378a^{5}r+653a^{4}r^{2}+514a^{3}r^{3}+142a^{2}r^{4}-35ar^{5}-25r^{6}\right)}{80ar^{2}(a+r)^{5}}\nonumber \\
 &  & -\frac{21\sqrt{\pi}a}{20r^{3}}\log\frac{a}{a+r},
\end{eqnarray}
and
\begin{eqnarray}
\psi_{1}^{(p)}(r) & = & \frac{\pi c_{1}^{2}\left(84a^{5}+378a^{4}r+658a^{3}r^{2}+539a^{2}r^{3}+192ar^{4}+15r^{5}\right)}{40r^{2}(a+r)^{5}}\nonumber \\
 &  & +\frac{21\pi ac_{1}^{2}}{10r^{3}}\log\frac{a}{r+a},\label{psi_1^p(r)}
\end{eqnarray}
from which one can find $\psi^{(p)}$, but we do not depict it here since it is a little
long.

Recall that for the ADM energy computation, we need the dominant terms
up to and including $\mathcal{O}(\frac{1}{r})$ in $\psi(r,\theta,\phi)$.
Collecting these parts in the above solutions, one gets

\begin{equation}
\psi(r)=1+\frac{a}{r}+\frac{J^{2}}{40a^{3}r}+\frac{5p^{2}}{32ar} +\mathcal{O}(\frac{1}{r^2}).\label{solution}
\end{equation}
Therefore from (\ref{asymp1}), the ADM energy of the solution reads

\begin{equation}
E_{\text{ADM}}=2a+\frac{J^{2}}{20a^{3}}+\frac{5p^{2}}{16a}.
\end{equation}
Observe that the $Jp$ term does not contribute to the energy since
it is of $\mathcal{O}(\frac{1}{r^{2}})$.

Next, as in \cite{Dennison-Baumgarte}, let us express the ADM energy
in terms of the irreducible mass $M_{\text{irr}}$ which is defined\cite{Chris}
as 
\begin{equation}
M_{\text{irr}}:=\sqrt{\frac{A}{16\pi}}\label{irreduciblemassformula}
\end{equation}
with $A$ being the area of a {\it section} of the event horizon. But as
the event horizon is a $4$ dimensional concept, which cannot be derived
from the $3$ dimensional initial data, we will approximate this with
the area of the apparent horizon, $A_{\text{AH}}$, following \cite{Dennison-Baumgarte}.

\section{{\normalsize{}{}Computation of the apparent horizon for the boosted,
rotating solutions}}

Let $S$ be a $2$ dimensional subspace of $\Sigma$ and $s^{i}$
be the normalized unit vector of $S$, {\it i.e.} $s^{i}s_{i}=1$. Then
the metric on $S$ is the pull-back metric from $\Sigma$ given as
\begin{equation}
m_{ij}:=\gamma_{ij}-s_{i}s_{j}.
\end{equation}
The expansion of the null geodesic congruence vanishes at the apparent
horizon by definition, \textit{i.e.} it is a marginally trapped surface
and the defining equation becomes 
\begin{equation}
\left(\gamma^{ij}-s^{i}s^{j}\right)\left(D_{i}s_{j}-K_{ij}\right)=0.\label{apparenthorizonequation1}
\end{equation}
Assuming the surface to be defined as a level set of a function 
\begin{equation}
\Phi:=r-h(\theta,\phi)=0,
\end{equation}
then the normal one-form reads 
\begin{equation}
s_{i}:=\lambda m_{i}=\lambda\partial_{i}\varPhi,
\end{equation}
which explicitly becomes 
\begin{equation}
s_{i}=\lambda\left(1,-\partial_{\theta}h,-\partial_{\phi}h\right).
\end{equation}
Recall that the metric on $\Sigma$ is 
\begin{equation}
\gamma_{ij}=\psi^{4}\begin{pmatrix}1 & 0 & 0\\
0 & r^{2} & 0\\
0 & 0 & r^{2}\sin^{2}\theta
\end{pmatrix},
\end{equation}
then one has

\begin{equation}
s^{i}=\lambda\left(\gamma^{rr},-\gamma^{\theta\theta}\partial_{\theta}h,-\gamma^{\phi\phi}\partial_{\phi}h\right),
\end{equation}
and 
\begin{equation}
\lambda=\Big(\gamma^{rr}+\gamma^{\theta\theta}(\partial_{\theta}h)^{2}+\gamma^{\phi\phi}(\partial_{\phi}h)^{2}\Big)^{-1/2}.
\end{equation}
Equation (\ref{apparenthorizonequation1}) reads more explicitly as

\begin{equation}
\gamma^{ij}\partial_{i}m_{j}-\gamma^{ij}\Gamma_{ij}^{k}m_{k}-\lambda^{2}m^{i}m^{j}\partial_{i}m_{j}+\lambda^{2}m^{i}m^{j}m_{k}\Gamma_{ij}^{k}+\lambda m^{i}m^{j}K_{ij}=0,\label{apparenthorizonequation}
\end{equation}
where we have used $\gamma^{ij}K_{ij}=K=0$. After working out each
piece, one arrives at 
\begin{eqnarray}
 &  & -\gamma^{\theta\theta}\partial_{\theta}^{2}h-\gamma^{\phi\phi}\partial_{\phi}^{2}h-\frac{1}{2}\Bigl((\gamma^{rr})^{2}\partial_{r}\gamma_{rr}-\gamma^{\theta\theta}\gamma^{rr}\partial_{r}\gamma_{\theta\theta}-\gamma^{\phi\phi}\gamma^{rr}\partial_{r}\gamma_{\phi\phi}+\partial_{\theta}h\gamma^{\phi\phi}\gamma^{\theta\theta}\partial_{\theta}\gamma_{\phi\phi}\Bigr)\nonumber \\
 &  & +\lambda^{2}\Bigl((\gamma^{\theta\theta})^{2}(\partial_{\theta}h)^{2}\partial_{\theta}^{2}h+(\gamma^{\phi\phi})^{2}(\partial_{\phi}h)^{2}\partial_{\phi}^{2}h+2\gamma^{\phi\phi}\gamma^{\theta\theta}\partial_{\phi}h\partial_{\theta}h\partial_{\theta}\partial_{\phi}h\Bigr)\nonumber \\
 &  & +\frac{\lambda^{2}}{2}\Bigl((\gamma^{rr})^{3}\partial_{r}\gamma_{rr}+(\gamma^{\theta\theta})^{2}\gamma^{rr}(\partial_{\theta}h)^{2}\partial_{r}\gamma_{\theta\theta}+(\gamma^{\phi\phi})^{2}\gamma^{rr}(\partial_{\phi}h)^{2}\partial_{r}\gamma_{\phi\phi}\nonumber \\
 &  & ~~~~~~~~~~~~~-(\partial_{\phi}h)^{2}\partial_{\theta}h(\gamma^{\phi\phi})^{2}\gamma^{\theta\theta}\partial_{\theta}\gamma_{\phi\phi}\Bigr)\nonumber \\
 &  & +\lambda\Bigl((\gamma^{rr})^{2}K_{rr}+(\gamma^{\theta\theta})^{2}(\partial_{\theta}h)^{2}K_{\theta\theta}+(\gamma^{\phi\phi})^{2}(\partial_{\phi}h)^{2}K_{\phi\phi}-2\gamma^{rr}\gamma^{\theta\theta}\partial_{\theta}hK_{r\theta}\nonumber \\
 &  & ~~~~~~~~~~~~~~-2\gamma^{rr}\gamma^{\phi\phi}\partial_{\phi}hK_{r\phi}+2\gamma^{\theta\theta}\gamma^{\phi\phi}\partial_{\theta}h\partial_{\phi}hK_{\theta\phi}\Bigr)=0.
\end{eqnarray}
An exact solution to this equation is beyond reach and we do not
really need it. All we need is an approximate solution of the form
\begin{equation}
h(\theta,\phi)=h^{0}+ph^{p}+Jh^{J}+\mathcal{O}(p^{2},J^{2},Jp),
\end{equation}
where 
\begin{equation}
\partial_{r}h=0,~~~~~~\partial_{r}h^{0}=0=\partial_{\theta}h^{0}=\partial_{\phi}h^{0}.
\end{equation}
Note that to compute the area of the apparent horizon and the irreducible mass up to and including the $\mathcal{O}(p^{2},J^{2},Jp)$ terms, one only needs the shape of the horizon up to and including  the $\mathcal{O}(p,J)$ terms which becomes clear when one studies the area integral. [See also\cite{Dennison-Baumgarte}.] 
Ignoring the higher order terms such as $(\partial_{\theta}h)^{2}$,
$(\partial_{\phi}h)^{2}$ and $\partial_{\theta}h\partial_{\phi}h$,
the apparent horizon equation becomes

\begin{align}
 & -\gamma^{\theta\theta}\partial_{\theta}^{2}h-\gamma^{\phi\phi}\partial_{\phi}^{2}h-\frac{1}{2}\left((\gamma^{rr})^{2}\partial_{r}\gamma_{rr}-\gamma^{\theta\theta}\gamma^{rr}\partial_{r}\gamma_{\theta\theta}-\gamma^{\phi\phi}\gamma^{rr}\partial_{r}\gamma_{\phi\phi}+\partial_{\theta}h\gamma^{\phi\phi}\gamma^{\theta\theta}\partial_{\theta}\gamma_{\phi\phi}\right)\nonumber \\
 & ~~~~~~~~~~~~~~~~~~~~+\frac{\lambda^{2}}{2}(\gamma^{rr})^{3}\partial_{r}\gamma_{rr}+\lambda\gamma^{rr}\left(\gamma^{rr}K_{rr}-2\gamma^{\theta\theta}\partial_{\theta}hK_{r\theta}-2\gamma^{\phi\phi}\partial_{\phi}hK_{r\phi}\right)=0.\label{apparenthorizonequation2}
\end{align}
To proceed, we need the components of the extrinsic curvature in the
$(r,\theta,\phi)$ coordinates. After coordinate transformations,
one finds 
\begin{equation}
\hat{K}_{rr}=\frac{3p}{r^{2}}\Big(c_{1}\sin\theta\cos\phi+c_{2}\cos\theta\Big),\,\,\,\hat{K}_{r\theta}=\frac{3p}{2r}\Big(c_{1}\cos\theta\cos\phi-c_{2}\sin\theta\Big),
\end{equation}
and 
\begin{equation}
\hat{K}_{r\phi}=-\frac{3p}{2r}c_{1}\sin\theta\sin\phi+\frac{3J}{r^{2}}\sin^{2}\theta.
\end{equation}
Therefore the resulting equation is 
\begin{equation}
\partial_{\theta}^{2}h+\frac{1}{\sin^{2}\theta}\partial_{\phi}^{2}h+\cot\theta\partial_{\theta}h-2r-4r^{2}\frac{\partial_{r}\psi}{\psi}+\frac{6J}{\psi^{4}r^{2}}\partial_{\phi}h-\frac{3p}{\psi^{4}}\Big(c_{1}\sin\theta\cos\phi+c_{2}\cos\theta\Big)=0.\label{result1}
\end{equation}
At order, $\mathcal{O}(p^{0},J^{0})$, this equation yields 
\begin{equation}
1+2r\frac{\partial_{r}\psi}{\psi}=0,
\end{equation}
where $\psi=1+\frac{a}{r}$ . And setting $r=h$, one finds 
\begin{equation}
h^{0}=a.
\end{equation}
This explains the physical meaning of the parameter $a$: it is the
location of the apparent horizon at the lowest order. The next order
contribution, which we shall find below, will be perturbations to
this location. At $\mathcal{O}(p)$ and $\mathcal{O}(J)$ we have
the following equations, respectively

\begin{equation}
\partial_{\theta}^{2}h^{p}+\frac{1}{\sin^{2}\theta}\partial_{\phi}^{2}h^{p}+\cot\theta\partial_{\theta}h^{p}-h^{p}-\frac{3}{16}\Big(c_{1}\sin\theta\cos\phi+c_{2}\cos\theta\Big)=0,
\end{equation}
and 
\begin{equation}
\partial_{\theta}^{2}h^{J}+\frac{1}{\sin^{2}\theta}\partial_{\phi}^{2}h^{J}+\cot\theta\partial_{\theta}h^{J}-h^{J}=0.
\end{equation}
These are linear PDE's and a close scrunity shows that $h^{J}$ equation
is the homogenous Helmholtz equation on a sphere ($S^{2}$), while
$h^{P}$ equation is the inhomogeneous Helmholtz equation with a non-trivial source.
So the next task is to find everywhere finite solutions of the following
equation 
\begin{equation}
\left({\vec{\nabla}}_{S^{2}}^{2}+k\right)f\left(\theta,\phi\right)=g\left(\theta,\phi\right),
\end{equation}
where ${\vec{\nabla}}_{S^{2}}^{2}$ is the Laplacian on $S^{2}$: 
\begin{equation}
{\vec{\nabla}}_{S^{2}}^{2}=\partial_{\theta}^{2}+\cot\theta\partial_{\theta}+\frac{1}{\sin^{2}\theta}\partial_{\phi}^{2}.
\end{equation}
It is clear that the Green's function technique is the most suitable
approach to this problem. For the Helmholtz operator on the sphere,
the Green function $G(\hat{x},\hat{x}')$ is defined as 
\begin{equation}
\left({\vec{\nabla}}_{S^{2}}^{2}+\lambda(\lambda+1)\right)G(\hat{x},\hat{x}')=
\delta^{(2)}(\hat{x}-\hat{x}'),
\end{equation}
which can be found to be (for example, see \cite{Green}) 
\begin{equation}
G(\hat{x},\hat{x}')=\frac{1}{4\sin\pi\lambda}\sum_{n=0}^{\infty}\frac{1}{\left(n!\right)^{2}}\frac{\Gamma(n-\lambda)}{\Gamma(-\lambda)}\frac{\Gamma(n+\lambda+1)}{\Gamma(\lambda+1)}\left(\frac{1+\hat{x}\cdot\hat{x'}}{2}\right)^{n},
\end{equation}
where $\hat{x}=\sin\theta\cos\phi\hat{i}+\sin\theta\sin\phi\hat{j}+\cos\theta\hat{k}$
and $\hat{x}^{'}$ is a similar expression with some other $\theta$
and $\phi$. Employing this Green's function with $\lambda=\frac{-1+i\sqrt{3}}{2}$
one finds

\begin{equation}
h^{p}=-\frac{1}{16}\left(c_{1}\sin\theta\cos\phi+c_{2}\cos\theta\right)
\end{equation}
and $h^{J}=0$. Therefore the apparent horizon is located at 
\begin{equation}
r=h(\theta,\phi)=a-\frac{1}{16}\Big(\vec{p}\cdot \hat{J}\cos\theta+\vert \vec{p}\wedge \hat{J}\rvert \sin\theta\cos\phi \Big),\label{hor5}
\end{equation}
where $\hat{J} = \frac{\vec{J}}{J}$. In the limit $\theta_{0}=0$, $h$
reduces to the form given in \cite{Dennison-Baumgarte}, that is 
$h(\theta)=a-\frac{p}{16}\cos\theta$; and the apparent horizon in this axially symmetric case is a squashed sphere from the North pole.  Note that the shape of the apparent horizon \ref{hor5} at this order does not depend on the magnitude of the spin but it does depend on its orientation with respect to the linear momentum. In figure 1, we plot the apparent horizon. To be able to see the dimple clearly in the whole figure, we have chosen a high momentum value.

\begin{figure}
\centering \includegraphics[width=0.4\linewidth]{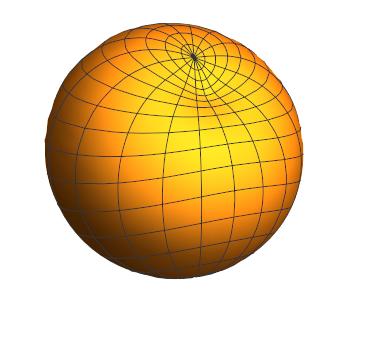} \caption{ The shape of the apparent horizon when the angle between $\vec{p}$ and
$\vec{J}$ is 45 degrees; and to be able to see the dimple, we have chosen $p/a=8\sqrt{2}$ which is outside the
validity of the approximation we have worked with. But the dimple exists for even small $p$.}
\label{fig:AH1} 
\end{figure}
Let us now evaluate the area of the apparent horizon from the formula
\begin{equation}
A_{\text{AH}}=\intop_{0}^{2\pi}d\phi\intop_{0}^{\pi}d\theta\sqrt{\det m},
\end{equation}
which at the order we are working yields 
\begin{equation}
A_{\text{AH}}=\intop_{0}^{2\pi}d\phi\intop_{0}^{\pi}d\theta\thinspace\sin\theta\thinspace\psi^{4}\thinspace h^{2}\left(1+\frac{1}{h^{2}}\left(\partial_{\theta}h\right)^{2}+\frac{1}{h^{2}\sin^{2}\theta}\left(\partial_{\phi}h\right)^{2}\right)^{1/2}.
\end{equation}
This is a pretty long computation since the conformal factor is quite complicated. But at the end, one finds   
\begin{equation}
A_{\text{AH}}=64\pi a^{2}+4\pi p^{2}+\frac{11\pi J^{2}}{5a^{2}}.
\end{equation}
Note that the angle between the spin and the linear momentum does not appear in the area. 
Then the irreducible mass $M_{\text{irr}}$ reads
\begin{equation}
M_{\text{irr}}=2a+\frac{p^{2}}{16a}+\frac{11J^{2}}{320a^{3}}.
\end{equation}
Comparing with $E_{ADM}$ we have
\begin{equation}
E_{\text{ADM}}=M_{\text{irr}}+\frac{p^{2}}{2M_{\text{irr}}}+\frac{J^{2}}{8M_{\text{irr}}^{3}}
\end{equation}
which matches the slow momentum and spin limit of the result in \cite{Chris}.
\section{Conclusions}
Momentum constraints in General Relativity are easily solved with the
method of Bowen-York while the Hamiltonian constraint is a nontrivial
elliptic equation. Here, extending earlier works \cite{Gleiser},
\cite{Dennison-Baumgarte} we gave an approximate analytical solution
that describes a spinning and moving system with a conserved spin
and linear momentum pointing in arbitrary directions. We computed
the properties of the apparent horizon, such as its shape and surface
area and showed the dependence of the shape on the angle between the
spin and the linear momentum. We calculated the relation between the
conserved quantities such as the ADM mass, the spin, the linear momentum
and the irreducible mass. The area of the apparent horizon does not depend on the angle between the spin and the linear momentum, but a dimple arises in the apparent horizon whose location depends on this angle.

\begin{acknowledgments}
We would like to thank Fethi Ramazano\u{g}lu for useful discussions.
\end{acknowledgments}

\end{document}